\documentclass[reprint,superscriptaddress,amsmath,amssymb,aps,pra,floatfix,showpacs]{revtex4-1}

\usepackage{graphicx}
\usepackage{dcolumn}
\usepackage{bm}
\usepackage{color}

\begin{document}


\title{Measurement of the Yb {\sc{I}} $^1S_0 - ^1P_1$ transition frequency at 399 nm using an optical frequency comb}

\author{Michaela Kleinert}
\affiliation{Department of Physics, Willamette University, Salem OR 97301, USA}
\author{M. E. Gold Dahl}
\affiliation{Department of Physics and Astronomy, Brigham Young University, Provo UT 84602, USA}
\author{Scott Bergeson}
 \email{scott.bergeson@byu.edu}
\affiliation{Department of Physics and Astronomy, Brigham Young University, Provo UT 84602, USA}

\date{\today}

\begin{abstract}
We determine the frequency of the Yb {\sc{I}} $^1S_0 - ^1P_1$ transition at 399 nm using an optical frequency comb. Although this transition was measured previously using an optical transfer cavity [D. Das et al., Phys. Rev. A \textbf{72}, 032506 (2005)], recent work has uncovered significant errors in that method. We compare our result of 751 526 533.49 $\pm$ 0.33 MHz for the Yb-174 isotope with those from the literature and discuss observed differences. We verify the correctness of our method by measuring the  frequencies of well-known transitions in Rb and Cs, and by demonstrating proper control of systematic errors in both laser metrology and atomic spectroscopy. We also demonstrate the effect of quantum interference due to hyperfine structure in a divalent atomic system and present isotope shift measurements for all stable isotopes.
\end{abstract}


\maketitle

\section{Introduction}

Optical frequency combs have revolutionized precision laser frequency measurements \cite{diddams00,udem02}. These combs make it possible to determine absolute laser frequencies across the visible \cite{Ye:00}, infrared \cite{schleisser12}, and ultraviolet wavelength ranges \cite{cignoz12} with an accuracy limited only by laboratory frequency standards \cite{heinecke2009}. Under ideal circumstances, the laser metrology is stable enough that relative  fractional frequency instabilities as low as $10^{-18}$ can be reliably measured \cite{hinkley2013}. The accuracy is great enough that frequency-comb based measurements of atomic transitions are being considered for the re-definition of the second \cite{Riehle2015506,targat13}.

Laser metrology methods based on frequency combs is more reliable than those based on wavelength measurements \cite{udem02}. This is due in part to the fact that time and frequency can be measured in the laboratory with greater reliability than distance. It is also due to the fact that frequency measurements are free from geometric distortions and phase shifts associated with wavelength measurements.

In this paper we report a measurement of the {Yb \sc{I}} $^1S_0 - ^1P_1$ transition frequency at 399 nm using an optical frequency comb. We verify the accuracy of our laser metrology by measuring the frequencies of several well-known transitions in Rb and Cs. We show how hyperfine interactions systematically shift the transition frequencies in the odd Yb isotopes, an effect not previously accounted for in Yb frequency measurements. Our measurements agree with the less accurate results of Refs. \cite{nizamani2010,Enomoto2016}. Our frequency comb measurements disagree with the value reported in Ref. \cite{das2005}, as discussed in the following section.

We also report measurements of the isotope shifts in the {Yb \sc{I}} $^1S_0 - ^1P_1$ transition. It could be argued from the standpoint of comparing with atomic structure calculations that isotope shift data is more important than absolute transition frequencies because the shifts can be calculated to higher accuracy than the absolute transition frequencies \cite{dzuba05, johnson94, lindgren84}.

\section{\label{sec:das} Previous measurements of the 399~\lowercase{nm} transition}

A determination of the {Yb \sc{I}} $^1S_0 - ^1P_1$ transition frequency was reported in Ref. \cite{das2005}. This measurement was based on a wavelength comparison between two lasers using an optical cavity. One laser at 798 nm was frequency doubled and used to probe the Yb transition. The other laser at 780 nm was stabilized to saturated absorption in Rb.

Optical cavities have been used to compare the wavelengths of widely separated laser lines with good accuracy.  This technique requires a careful measurement of the cavity's free spectral range as well as its phase relation with wavelength (see, for example \cite{Sterr:95} and \cite{bergeson98} footnote 14). In its most successful implementation, this method has produced sub-MHz accuracy with results that have been reproduced by different research groups \cite{BARR1985217,Gillaspy:91,Gillaspy:92}.

The optical cavity method used in Ref. \cite{das2005} differs from previous work in that a bow-tie cavity was used instead of a linear one. This method has produced good measurement results when the transition frequency was previously known with high enough accuracy. For example, a recent measurement of the Cs-133 D1 transitions by this group has reproduced the results of earlier high-accuracy frequency comb measurements \cite{Singh:12}.

However, when the transition frequency is not well known, the method of Ref. \cite{das2005} produces unreliable results \cite{falke2006}. The K-39 D1 and D2 hyperfine-free transitions published by this group disagree strongly with frequency comb measurements by 478 MHz and 592 MHz, respectively \cite{falke2006}, even though the uncertainties were estimated to be 0.05 and 0.1 MHz. The authors of the frequency comb work concluded that the initial frequency used in the optical cavity method were ``not sufficiently precise to unambiguously determine the D lines’ frequencies.'' \cite{falke2006}

Even when the transition frequencies are known well enough to give reliable laser metrology, the spectroscopic methods used by this group have systematic errors larger than anticipated in some cases. For example, the Li-6 D2 $F=\textstyle{\frac{1}{2}}\rightarrow F^{\prime}=\textstyle{\frac{3}{2}}$ transition frequency measured by this group disagrees with the frequency comb measurements of Ref. \cite{sansonetti2011} by 1.85 MHz and the Li-7 $F=2\rightarrow F^{\prime}=3$ transition by 0.65 MHz even though the uncertainties were though to be only 0.060 and 0.040 MHz, respectively (31 and 16 standard deviations). In these measurements, the influence of the hyperfine interaction and the variation in the apparent transition frequency with laser polarization was not considered. This omission alone leads to MHz-level systematic errors \cite{brown2013}. Similarly, MHz-level discrepancies with the Rb D1 transitions from this group are found when they are compared with the frequency comb work of Ref. \cite{maric2008}. These discrepancies need to be considered in context. In some cases the results from this group are truly impressive \cite{das2007,das2008}.

Previous to the measurements in Ref. \cite{das2005}, only the NIST Atomic Spectra Database data was available for the absolute frequency of the Yb I 399 nm transition  \cite{NIST_ASD}. The uncertainties related to these data are perhaps $\pm 150$ MHz, although the uncertainties are not well characterized \cite{kramida2016}. For the reasons given above, the measurements of this Yb transition in Ref. \cite{das2005} must be considered with caution.

Somewhat more recently, two other measurements of the 399 nm transition have been published. The measurement in Ref. \cite{nizamani2010} used a wavemeter to determine the absolute transition frequency in Yb-176. The accuracy of the measurement was limited by the 60 MHz absolute accuracy of the wavemeter. Another measurement in Ref. \cite{Enomoto2016} used an optical cavity to span a 41 THz optical frequency gap between a known laser frequency and a probe laser. This method is similar to that used by Ref. \cite{das2005}, and the estimated uncertainty is 100 MHz. Given these data and the unreliability of the measurement in Ref. \cite{das2005}, a new determination of the Yb I 399 nm transition frequency is warranted.

\section{The laser system}

\begin{figure} \centerline{\includegraphics[width=\columnwidth]{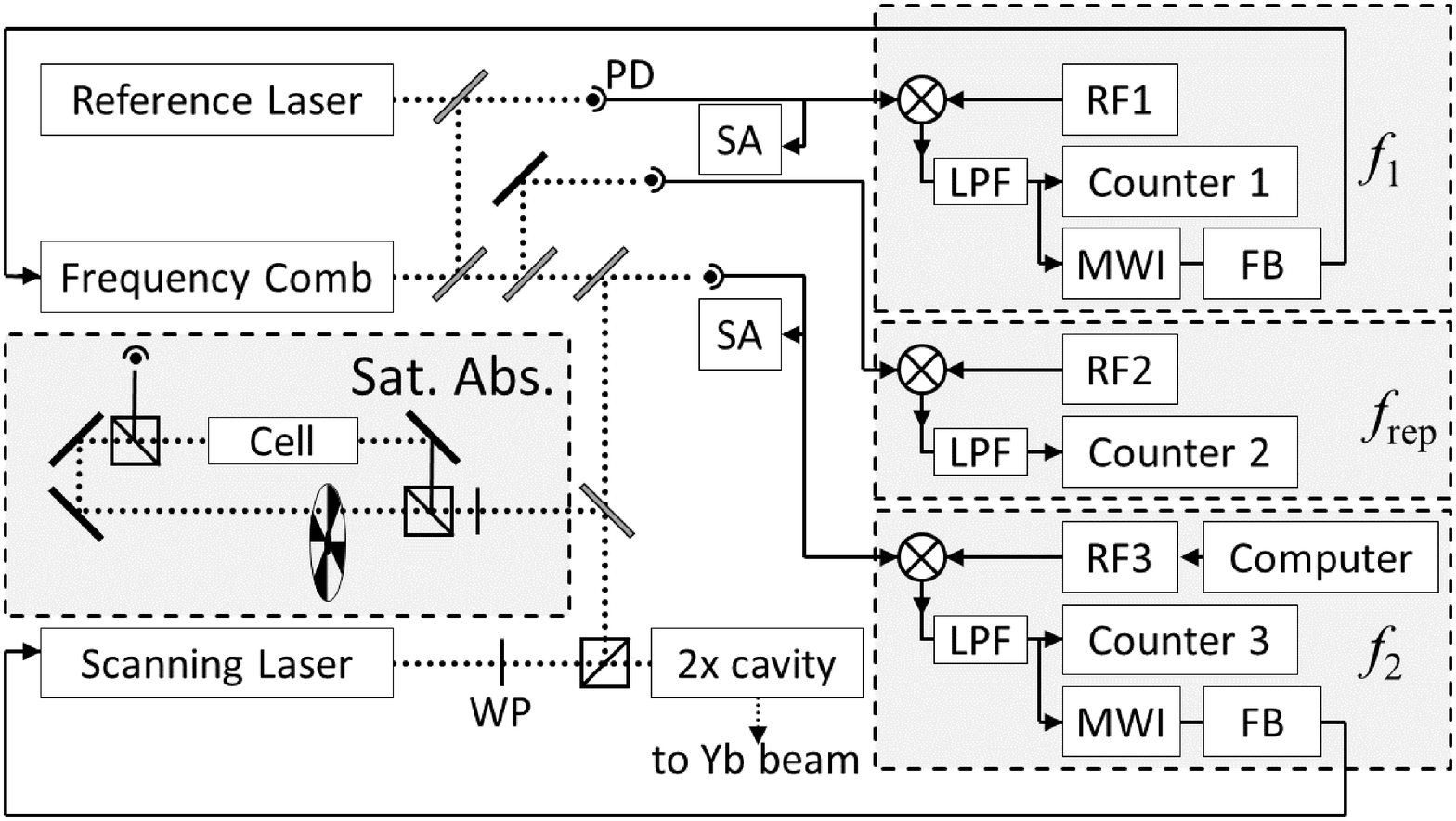}}
  \caption{\label{fig:layout}
  Schematic diagram of the laser system. The frequency comb is a femtosecond laser with a 984 MHz repetition rate. The reference laser is a diode laser locked to the Rb D2 $F=2\rightarrow F^{\prime}=(2,3)$ crossover transition. The beatnote frequency $f_1$  between the reference laser and the frequency comb is stabilized using a microwave interferometer and feedback control. The scanning laser is a Ti:Sapphire laser. The beatnote frequency $f_2$ between the scanning laser and comb is also stabilized. The accuracy of the offset-locking scheme is evaluated by  measuring transition frequencies in Cs and Rb using saturated absorption (Sat. Abs.). PD = photodiode, SA = spectrum analyzer, LPF = low pass filter, RF1,2,3 = radio frequency synthesizers, MWI = microwave interferometer, FB = feedback control, WP = waveplate.
  }
\end{figure}

A schematic diagram of the laser system is shown in Fig. \ref{fig:layout} \cite{Lyon:14}. The frequency comb is generated by a femtosecond laser oscillator (Laser Quantum Gigajet) with a repetition rate $f_{\rm rep} \approx$ 984 MHz. The repetition rate is measured using a high-speed photodiode. By mixing the photodiode signal down to 9 MHz using a stable RF synthesizer, we measure $f_{\rm rep}$ with a precision of 0.1 Hz.

The reference laser in Fig. \ref{fig:layout} is a diode laser (Vescent Photonics DFB Laser Module) that is locked to the Rb-87 D2 $F=2 \rightarrow F^{\prime} = (2,3)$ crossover transition near 780 nm. The beatnote between the reference laser and the nearest comb mode is locked to a particular value using a microwave interferometer and feedback control. The beatnote frequency $f_1$, which is measured directly using a spectrum analyzer, is combined with a stable RF signal and mixed down to 20 MHz. This filtered and amplified signal is sent to the microwave interferometer consisting of a power splitter, a delay line, a frequency mixer, and a low-pass filter. The output of the interferometer is a low-noise dispersion-like dc-signal that we use to offset-lock the nearest comb mode to the reference laser. This signal feeds back to the frequency comb's cavity length \cite{Walker2007}. Therefore, the frequency of one mode of the comb is well-known, provided that the saturated absorption lock in the reference laser is accurate.Counting $f_{\rm rep}$ then gives us the absolute frequencies of all of the other modes in the comb.

The scanning laser in Fig. \ref{fig:layout} is a Ti:Sapphire laser (M-Squared Lasers SolsTiS). A portion of the laser beam is split off and referenced to the frequency comb in a manner that is similar to the reference laser, producing the beatnote $f_2$. The only difference is that the RF synthesizer (RF3 in Fig. \ref{fig:layout}) is controlled by a computer.

The frequency interval, $df$, between the reference laser and the scanning laser is given by
\begin{equation}
  df = n f_{\rm rep} \pm f_1 \pm f_2,
  \label{eqn:df}
\end{equation}
where $n$ is an integer. The ambiguity of the signs in Eq. \ref{eqn:df} is resolved by experimentally observing how the magnitudes of the beatnote frequencies change as the entire comb shifts up and down in frequency. This shift is accomplished by varying the frequency comb cavity length slightly. The absolute frequency of the scanning laser, $f_{\rm SL}$ is then given by
\begin{equation}
  f_{\rm SL} = f_{\rm Rb} - df,
  \label{eqn:fscan}
\end{equation}
where $f_{\rm Rb}$ is the frequency of the Rb-87 D2 $F=2 \rightarrow F^{\prime}=(2,3)$ crossover transition \cite{Ye:96}. The integer $n$ in Eqs. (\ref{eqn:df}) and (\ref{eqn:fscan}) is reduced to $\leq\pm1$ by measuring the wavelength of the scanning laser. We use a Toptica High Finesse WA-6 wavemeter with an absolute accuracy of 600 MHz. The final ambiguity in $n$ is eliminated by measuring the Yb transition frequency for different values of $f_{\rm rep}$.

\section{Accuracy of the frequency comb}
\label{sec:accuracy}

Accuracy issues generally divide into two categories. One is laser metrology, or the ability to measure laser frequencies correctly. In our experiment, the frequency counters, general counting errors in the beatnotes, reference laser lock errors, and frequency comb errors contribute to this category. The second category is atomic spectroscopy, or the ability to accurately interrogate the atomic transitions. These issues include Zeeman shifts, hyperfine and laser polarization shifts, laser-power related errors including Stark shifts, cell shifts, line-shape errors, and first-order Doppler shifts.

\begin{table}
\setlength{\tabcolsep}{8pt}
\caption{\label{tab:rbtransitions} Transitions in the Rb-87 D2 array used to determine the value of the reference laser offset lock. The known transition frequencies $f_0^{}$ were taken from Ref. \cite{Ye:96} and the convenient tables of Ref. \cite{steck}. The number in parenthesis in column 2 indicate the 1$\sigma$ standard deviation in repeated measurements of the transition frequency. These transitions were chosen because they are relatively insensitive to variations in laser power and polarization \cite{Grimm1989, Schmidt1994}.}
\begin{tabular}{lcc} \hline
Transition & $f-f_0^{}$ (MHz) & FWHM (MHz) \\ \hline \hline
$F=2 \rightarrow F^{\prime}=2,3$ & -0.033(6) & 7.57 \\
$F=2 \rightarrow F^{\prime}=1,3$ & -0.059(9) & 6.90 \\
$F=2 \rightarrow F^{\prime}=1,2$ & -0.059(5) & 7.42 \\
\hline
Mean & -0.050(15) & \\ \hline \hline
\end{tabular}
\end{table}

\subsection{Frequency counters and synthesizers}

All of the counters and RF synthesizers are referenced to a GPS-disciplined 10 MHz frequency standard with an absolute accuracy of $\Delta f/f = 1.6 \times 10^{-12}$. We use a Trimble Bullet antenna and Thunderbolt E GPS Disciplined Clock, to which we lock a Stanford Research Systems FS725 frequency standard.

If the beatnote signals $f_1$ and $f_2$ are noisy or weak they will not be counted properly. For all of the measurements reported here, the beatnotes are typically greater than 25 dB above the noise floor, measured using a 30 kHz resolution bandwidth. We compare the frequency of the RF synthesizers, the counters, and the spectrum analyzers and find that the counters accurately count the mixed-down beatnotes $f_1$ and $f_2$ with an error less than 20 kHz.

\subsection{Reference laser lock offset}
The reference laser is locked to the Rb-87 D2 $F=2\rightarrow F^{\prime}=(2,3)$ crossover transition using saturated absorption spectroscopy. The locked laser frequency depends on the zero-crossing in the error signal, which in turn depends on a dc-offset voltage in the lock circuit. Any errors in the dc-offset voltage translate directly into a frequency offset of the laser relative to the actual line center.

We determine the reference laser lock frequency offset by measuring the Rb-87 D2 transitions listed in Table \ref{tab:rbtransitions} with the scanning laser in a standard saturated-absorption experiment (see Fig. \ref{fig:layout}). Counter-propagating orthogonally-polarized laser beams overlap in a Rb reference cell (Triad technology TT-RB-75-V-P, 3 inches long and 1 inch diameter) that is at room temperature and surrounded by a double layer of Mu-metal to shield it from ambient magnetic fields. The pump to probe power ratio is close to 4. The Gaussian beam waist is 3.1 mm. The pump beam intensity is 0.25 mW/cm$^2$. A chopper wheel modulates the pump beam at a frequency of 770 Hz and a lock-in amplifier is used to extract the saturated absorption signal with a signal-to-noise ratio of a few hundred.

The data in Table \ref{tab:rbtransitions} indicate that our reference laser is locked $0.050 \pm 0.015$ MHz below the known Rb transition frequency, where the uncertainty is the 1$\sigma$ standard deviation of the un-weighted mean of column 2 in Table \ref{tab:rbtransitions}. Day-to-day reproducibility is better than $\pm 0.015$ MHz on the individual lines. These particular transitions were chosen because they are insensitive to changes in the laser polarization and to the pump and probe laser beam intensities.

\subsection{Frequency comb errors}

\begin{table}
\caption{\label{tab:transitions}
  Reference transitions used to demonstrate the accuracy of the comb in measuring frequency intervals up to 30,000 GHz. The number in parenthesis represents the 1$\sigma$ standard deviation in repeated measurements. The known transition frequencies $f_0$ are taken from Ref. \cite{gerginov2004} for Cs and Ref. \cite{maric2008} for Rb-87 D1. The integer $n$ from Eq. (\ref{eqn:df}) is shown in the last column. The 0.050 MHz correction from Table \ref{tab:rbtransitions} has been applied to these data, and the uncertainty in the mean at the bottom of the table is the 1$\sigma$ standard deviation of the numbers in column 2 added in quadrature with the 0.015 MHz uncertainty from Table \ref{tab:rbtransitions}.
}
\begin{tabular}{lcc} \hline
Transition & $f_{\rm SL} - f_0$ (MHz) & $n$ \\ \hline \hline
Cs D2 $F=3 \rightarrow F^{\prime}=3,4$    &  0.027(5)  & 33,004 \\
Cs D2 $F=3 \rightarrow F^{\prime}=2,3$    &  0.067(15) & 33,004 \\
Cs D2 $F=4 \rightarrow F^{\prime}=4,5$    &  0.053(10) & 33,013 \\
Cs D2 $F=4 \rightarrow F^{\prime}=2,3$    &  0.008(7)  & 33,013 \\
Rb-87 D1 $F=2 \rightarrow F^{\prime}=1,2$ & -0.029(22) & 8,597 \\
\hline
Mean & 0.024(40) & \\ \hline \hline
\end{tabular}
\end{table}

In a fully stabilized frequency comb, both the pump laser power and the cavity length are controlled. The cavity length is often used to stabilize $f_{\rm rep}$ and the pump laser power is used to control the carrier-envelope offset phase, although other configurations are possible \cite{Walker2007}. Because the pump power influences both the carrier-envelope offset phase and the cavity length, these control parameters are not independent.

In our experiment we only control the cavity length and we use it to stabilize the frequency of only one cavity mode. This is similar to the method described in Section V.A of Ref. \cite{heinecke2009}, except that the frequency comb repetition rate in our work is only counted, not stabilized.

\begin{figure}
  \centerline{\includegraphics[width=\columnwidth]{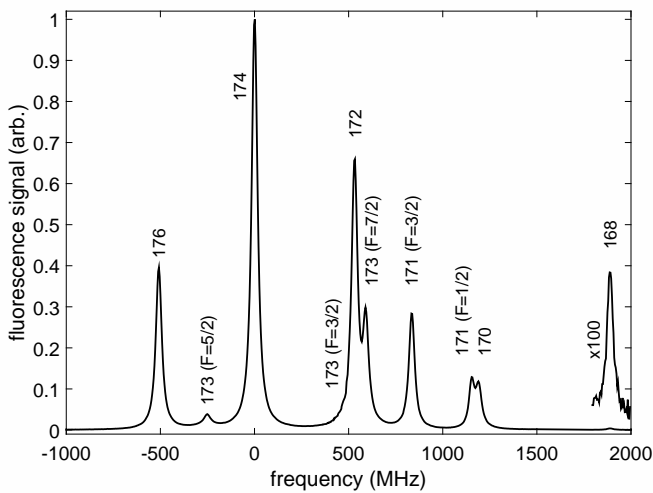}}
  \caption{\label{fig:ybscan} Laser-induced fluorescence measurements of Yb atoms in the atomic beam. Zero frequency corresponds to the center of the Yb-174 transition. The different isotopes and upper state hyperfine levels are labeled in the plot.}
\end{figure}

In this ``partially-stabilized'' configuration \cite{Lyon:14}, we find that the uncontrolled repetition rate varies by approximately 1 Hz in repeated 1-second measurements. This level of variation is negligible in our experiment because the largest frequency interval that we measure is between the Rb D2 transition at 780 nm and Cs D2 transition at 852 nm, corresponding to 30,000 GHz or 30,000 comb modes. The 1-Hz variation in the repetition rate contributes only 30 kHz of variability in this laser frequency interval. However, as we have shown previously \cite{Lyon:14}, even this variation is dramatically suppressed when all of the counters are read simultaneously, as we do in our experiment. The benefit of operating the comb in this partially-stabilized way is that the comb runs reliably without intervention all day long.

To verify our ability to count frequency intervals correctly, we measure a few well-known transitions in Cs and Rb, as shown in Table \ref{tab:transitions}. These data show that the frequency comb reliably measures frequency intervals as large as 30,000 GHz with 0.04 MHz accuracy. This uncertainty is a distributed error and indicates the errors related to our saturated absorption measurements as well as all of the laser metrology systematic errors.

\subsection{Cell shifts}

While it is straightforward to perform saturated absorption spectroscopy in Rb and Cs, the accuracy of such measurements is an issue even when the laser metrology is perfect. Frequency shifts specific to a given absorption cell can be surprisingly large. This issue was recently treated in depth \cite{Wu:13,PhysRevA.92.042504}, and it was shown that cell-shifts as large as 400 kHz can exist. These shifts, when present, can be estimated by measurements of the linewidth of the atomic transitions at low laser power. Our narrowest  lines for the Rb D2 transition are 6.9 MHz (FWHM), somewhat larger than the known value of 6.1 MHz. Similarly, in Cs our measured linewidths are 6.0 to 6.2 MHz, somewhat larger than the known value of 5.2 MHz. Given the analysis in Refs. \cite{PhysRevA.92.042504} and \cite{Wu:13}, these data suggest shifts of perhaps 20 kHz, although this is in reality just an estimate.

\section{Yb measurements}

\begin{figure}
  \centerline{\includegraphics[width=0.6\columnwidth]{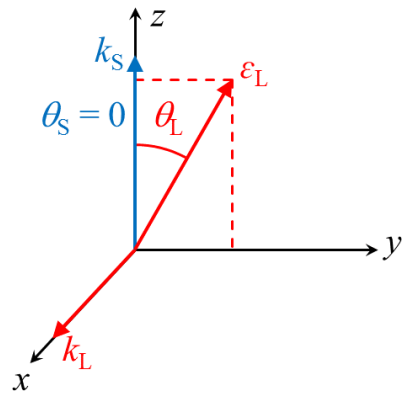}}
  \caption{\label{fig:beamconfig} (Color online.) A representation of the beam geometry used in our Yb measurements, following the notation of Ref. \cite{brown2013}. The laser beam travels in the $x$-direction, represented by the red arrow labeled $k_{\rm L}$. The scattered fluorescence signal is collected in the direction labeled by the blue arrow $k_{\rm S}$. The polarization of the laser is represented by the red arrow labeled $\epsilon_{\rm L}$. The angle between the laser polarization and vector pointing to the detector is represented by $\theta_{\rm L}$.
}
\end{figure}

We measure the Yb transition frequencies using laser-induced fluorescence in a collimated atomic beam of Yb. The scanning laser at a wavelength of 798 nm is frequency-doubled to 399 nm, and its frequency is controlled by a computer. To check for systematic effects, the 399 nm laser beam alternately crosses the Yb atomic beam at a right angle in two different directions. The laser beam has an intensity of $14~\mbox{mW/cm}^2 = I_{\rm sat}/4$, where $I_{\rm sat} = 57~\mbox{mW/cm}^2$ is the saturation intensity. The laser beam is retro-reflected with an angular error of approximately $\pm 0.2$ mRad. Fluorescent laser light is collected by an $f/\#=2$ achromat lens and measured using a photomultiplier tube (PMT) and an oscilloscope. Data is then transferred to a computer for analysis. A composite scan across the Yb $6s^2~^1S_0 - 6s6p~^1P_1$ transition for all isotopes is shown in Fig. \ref{fig:ybscan}.

\begin{figure}
  \centerline{\includegraphics[width=\columnwidth]{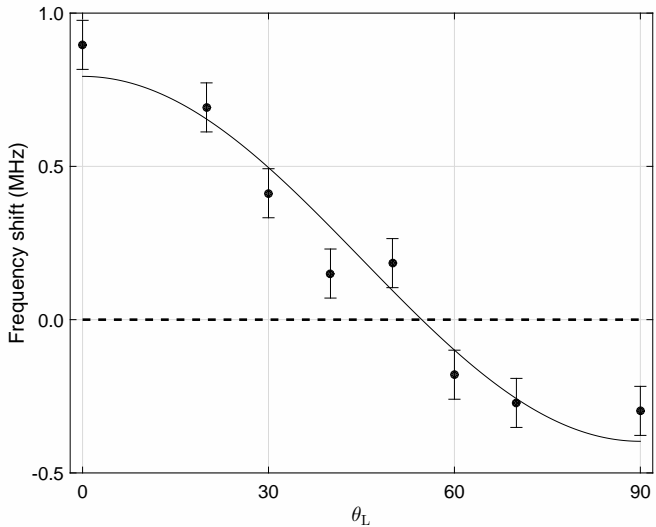}}
  \caption{\label{fig:angle} The change in transition frequency for the Yb-171 ($F=3/2$) transition as a function of the angle $\theta_{\rm L}$ for the geometry shown in Fig. \ref{fig:beamconfig}(a). The black circles are the measured data, the thin black line is the model of Eq. (\ref{eqn:angle}) with ${\cal{B}}=0.764$ MHz, and the dashed line indicates the expected value when $\theta_{\rm L}=54.73^{\circ}$. The errorbars indicate a typical 1$\sigma$ standard deviation in repeated measurements of the transition frequency in this measurement set.
  }
\end{figure}

The Yb atomic beam design is similar to Refs. \cite{senaratne2015} and \cite{guttridge2016}. A V-shaped hole is milled into a 2.75 inch double-sided conflat flange. This hole is filled with 136 microcapillaries with dimensions of 8 mm length, 0.2 mm inner diameter, and 0.4 mm outer diameter. A zero-length reducer flange is used to connect this flange to a 0.75 inch diameter 2.0 inch long weld stub on a 1.33 inch conflat flange, into which 25 grams of Yb metal is placed. The capillary flange is heated to 530 $^{\circ}$C, causing the Yb reservoir to reach a temperature of 434 $^{\circ}$C. A 6 mm diameter collimating aperture is located approximately 200 mm downstream of the capillary flange. At the location of this collimating aperture, we measure 6\% absorption of a weak probe beam that passes through the atomic beam. The atomic beam passes through a 40 cm long tube into the experimental chamber where the laser-induced fluorescence measurements are made. The end of this tube is a 16 mm diameter aperture for the atomic beam. The resulting atomic beam is uniform, with well-defined boundaries, having a divergence of 14 mRad. Using three orthogonal sets of large Helmholtz coils and a milliGauss probe, we zero the magnetic field in the center of our vacuum chamber to less than $\pm 0.03$ G. The Zeeman shift associated with this residual field is approximately $\pm 0.05$ MHz.

\subsection{\label{sec:hyperfine} Hyperfine structure}

The Yb-171 and Yb-173 isotopes have hyperfine structure. Quantum interference that arises from this structure can adversely influence the determination of the transition frequencies if not properly addressed \cite{brown2013,marsman2014,yost2014,amaro2015a,amaro2015b,beyer15}. The magnitude of this interference effect depends on laser intensity and polarization as well as measurement geometry. This effect has been measured in Li \cite{brown2013} and H \cite{beyer15}. It has been estimated in muonic hydrogenic atoms \cite{amaro2015a} and microwave transitions in He \cite{marsman2014}. However it has not been previously measured in divalent atoms. Understanding and controlling this effect is particularly important because isotope shift spectroscopy is used not only to study the structure of the nucleus \cite{Campbell2016127,ruiz2016} but also as a probe for physics beyond the standard model \cite{frugiuele2016,delaunay2016}.

Following the treatment in Ref. \cite{brown2013}, we will define our measurement geometry as shown in Fig. \ref{fig:beamconfig}. The $x$-direction is taken as the laser beam propagation direction, $k_{\rm L}$. The polarization of the laser beam, $\varepsilon_{\rm L}$, therefore lies in the $yz$-plane, and makes an angle $\theta_{\rm L}$ with respect to the $z$-axis. The PMT detector lies along the $z$-axis. It detects scattered light in a direction $k_{\rm S}$. There is no polarizer in the detection channel.

Quantum interference effects shift the hyperfine levels. The interaction energy can be written parametrically as \cite{brown2013}
\begin{equation}
 E_i = {\cal{A}} + \frac{{\cal{B}}}{2}\left(3 \cos^2\theta_{\rm S} \cos^2\theta_{\rm L} -1 \right),
 \label{eqn:angle}
\end{equation}
where $E_i$ represents the energy of level $i$, and the angles $\theta_{\rm S}$ and $\theta_{\rm L}$ have been defined above (see Fig. \ref{fig:beamconfig}). The parameters ${\cal{A}}$ and ${\cal{B}}$ depend on the transition line strengths and the cross-term interference. In the configuration shown in Fig. \ref{fig:beamconfig}, the angular dependence vanishes when $\theta_{\rm L} = 54.73^{\circ}$.

We have measured the apparent transition frequencies as a function of the laser polarization angle. The results are plotted in Fig. \ref{fig:angle}. The angle-dependence is significant, and if not properly treated can result in a MHz-level systematic error in determining the line center. For all of the measurements reported here, we use the geometry shown in Fig. \ref{fig:beamconfig} with $\theta_{\rm L} \approx 54.73^{\circ}$. Note that this polarization issue was not addressed in the measurements of Ref. \cite{das2005}.

\subsection{Isotope shifts}

\begin{table}[t!]
\caption{\label{tab:isotope}
Measured isotope shifts for the Yb {\sc{I}} 399 nm transition. The listed frequencies $f$ are given relative to Yb-174. The early work of Ref. \cite{chaiko1966} has not been included because it is significantly different from all recent measurements. The third column shows the difference between this work and previous measurements. The fourth column shows the magnitude of that difference in units of the combined uncertainties in the measurements. In column 1, the notation 173.52 refers to the $F=5/2$ hyperfine level in Yb-173, etc.
}
\begin{tabular}{llrcc}
Isotope	& $f$ (MHz) & $\Delta f$ & $\Delta f/\sigma$ & Ref. (year) \\ \hline \hline
176		& $-508.89	\pm	0.09 $&	$\cdots$&$\cdots$& This	work	\\
		& $-509.310	\pm	0.050$&	0.42	& 4.1	& \cite{das2005} (2005)	\\	
		& $-509.98	\pm	0.75 $&	1.09	& 1.4	& \cite{banerjee2003} (2003)	\\	
		& $-507.2	\pm	2.5	 $&	-1.69	& 0.7	& \cite{loftus2001} (2001)	\\	
		& $-509.4	\pm	4.0	 $&	0.51	& 0.1	& \cite{Grundevik1979} (1979) \\	
		& $-509		\pm	30	 $&	0.11	& 0.0	& \cite{nizamani2010} (2010) \\	
173.52	& $-250.78	\pm	0.33 $&	$\cdots$&$\cdots$& This	work	\\
		& $-253.418	\pm	0.050$&	2.64	& 7.9	& \cite{das2005} (2005)	\\	
		& $-254.67	\pm	0.63 $&	3.89	& 5.5	& \cite{banerjee2003} (2003) \\	
		& $-264		\pm	30	 $&	13.22	& 0.4	& \cite{nizamani2010} (2010) \\	
172		& $531.11	\pm	0.09 $&	$\cdots$&$\cdots$& This	work	\\
		& $533.309	\pm	0.053$&	-2.20	& 21.1	& \cite{das2005} (2005)	\\	
		& $533.90	\pm	0.70 $&	-2.79	& 4.0	& \cite{banerjee2003} (2003) \\	
		& $527.8	\pm	2.8	 $&	3.31	& 1.2	& \cite{loftus2001} (2001)	\\	
		& $529.9	\pm	4.0	 $&	1.21	& 0.3	& \cite{Grundevik1979} (1979) \\	
		& $546		\pm	60	 $&	-14.89	& 0.2	& \cite{nizamani2010} (2010) \\	
173.72	& $589.75	\pm	0.24 $&	$\cdots$&$\cdots$& This	work	\\
		& $587.986	\pm	0.056$&	1.76	& 7.2	& \cite{das2005} (2005)	\\	
		& $589.00	\pm	0.45 $&	0.75	& 1.5	& \cite{banerjee2003} (2003) \\	
		& $578.1	\pm	5.8	 $&	11.65	& 2.0	& \cite{loftus2001} (2001)	\\	
		& $546		\pm	60	 $&	43.75	& 0.7	& \cite{nizamani2010} (2010) \\	
171.32	& $835.19	\pm	0.20 $&	$\cdots$&$\cdots$& This	work	\\
		& $832.436	\pm	0.050$&	2.75	& 13.4	& \cite{das2005} (2005)	\\	
		& $833.24	\pm	0.75 $&	1.95	& 2.5	& \cite{banerjee2003} (2003) \\	
		& $832.5	\pm	5.6	 $&	2.69	& 0.5	& \cite{loftus2001} (2001)	\\	
		& $834.4	\pm	4.0	 $&	0.79	& 0.2	& \cite{Deilamian:93} (1993) \\	
		& $829		\pm	30	 $&	6.19	& 0.2	& \cite{nizamani2010} (2010) \\	
171.12	& $1153.68	\pm	0.25 $&	$\cdots$&$\cdots$& This	work	\\
		& $1153.696	\pm	0.061$&	-0.02	& 0.1	& \cite{das2005} (2005)	\\	
		& $1152.86	\pm	0.60 $&	0.82	& 1.3	& \cite{banerjee2003} (2003) \\	
		& $1151.4	\pm	5.6	 $&	2.28	& 0.4	& \cite{loftus2001} (2001)	\\	
		& $1135.2	\pm	5.8	 $&	18.48	& 3.2	& \cite{Deilamian:93} (1993) \\	
		& $1149		\pm	60	 $&	4.68	& 0.1	& \cite{nizamani2010} (2010) \\	
170		& $1190.36	\pm	0.49 $&	$\cdots$&$\cdots$& This	work	\\
		& $1192.393	\pm	0.055$&	-2.03	& 4.1	& \cite{das2005} (2005)	\\	
		& $1192.48	\pm	0.9	 $&	-2.12	& 2.1	& \cite{banerjee2003} (2003) \\	
		& $1172.5	\pm	5.7	 $&	17.86	& 3.1	& \cite{Deilamian:93} (1993) \\	
		& $1175.7	\pm	8.1	 $&	14.66	& 1.8	& \cite{loftus2001} (2001)	\\	
		& $1195.0	\pm	10.8 $&	-4.64	& 0.4	& \cite{Grundevik1979} (1979) \\	
		& $1149		\pm	60	 $&	41.36	& 0.7	& \cite{nizamani2010} (2010) \\	
168		& $1888.80	\pm	0.11 $&	$\cdots$&$\cdots$& This	work	\\
		& $1887.400	\pm	0.050$&	1.40	& 11.6	& \cite{das2005} (2005)	\\	
		& $1886.57	\pm	1.00 $&	2.23	& 2.2	& \cite{banerjee2003} (2003) \\	
		& $1870.2	\pm	5.2	 $&	18.6	& 3.6	& \cite{Deilamian:93} (1993) \\	
		& $1883		\pm	30	 $&	5.80	& 0.2	& \cite{nizamani2010} (2010) \\	\hline
\end{tabular}
\end{table}

Isotope shifts in the 399 nm transition are listed in Table \ref{tab:isotope} relative to the Yb-174 isotope. Because we are measuring a difference between transition frequencies, many systematic errors subtract out. The data in Table \ref{tab:isotope} are influenced by the metrology errors, which are less than 0.04 MHz (see Sec. \ref{sec:accuracy}). The uncertainty in the measurements also includes errors related to the atomic spectroscopy. The statistical uncertainty due to fitting the lines is typically 0.04 MHz for the strongest lines, estimated from both variation in  repeated measurements and from models of the Voigt lineshape fitting process. The signal-to-noise ratio is greater than 85 for all the transitions reported here. The measured Lorentzian linewidth (FWHM) of the transitions is typically 33 MHz,  close to the natural linewidth of 28 MHz. Some details of the signal-to-noise ratios, measured linewidths, and statistical fit errors are given in Table \ref{tab:snr}.

\begin{table}
\caption{Signal-to-noise (SNR) ratios and full-widths at half-maximum (FWHM) for the transitions reported in this paper. The SNR is calculated as the height of a peak divided by the standard deviation of the fit residuals. The value given is the mean of a series of repeated measurements at a given laser power and alignment. The fit error is the 1$\sigma$ standard deviation of the fitted line center in repeated measurements at a given laser power and alignment. The natural width of the Yb transition is 28 MHz. \label{tab:snr}}
\begin{tabular}{lrcc}
Isotope & SNR & FWHM (MHz) & Fit error (MHz) \\ \hline \hline
176         & 450 & 35.2 & 0.03 \\
173 (F=5/2)	& 250 & 42.1 & 0.02 \\
174         & 560 & 30.2 & 0.02 \\
172	        & 190 & 34.0 & 0.02 \\
173 (F=7/2)	&  85 & 30.5 & 0.03 \\
171 (F=3/2)	& 870 & 30.8 & 0.02 \\
171 (F=1/2) & 250 & 31.8 & 0.05 \\
170	        & 150 & 33.6 & 0.02 \\
168         & 170 & 29.9 & 0.05 \\ \hline
\end{tabular}
\end{table}

\textbf{Yb-176, Yb-172, Yb-170, and Yb-168:} The 1$\sigma$ standard deviation in repeated measurements of the isotope shifts for Yb-176 and Yb-172 using different day-to-day laser alignment and different measurement configurations is 0.03 MHz. It is slightly larger, 0.08 MHz, for Yb-168 and much larger, 0.49 MHz, for Yb-170. To these we add the 0.08 MHz laser metrology uncertainty, twice the value in Table \ref{tab:transitions} because the laser is frequency doubled, in quadrature to obtain the estimated 1$\sigma$ uncertainties listed in Table \ref{tab:isotope}. Our data disagree with the data of Ref. \cite{das2005} for isotopes 176, 172, 170, and 168 by 0.42, -1.20, -2.03, and 1.40 MHz, each by several combined standard deviations. This is a level similar to the variation seen in comparisons with other measurements from this group with frequency comb measurements.

\textbf{Yb-171 and Yb-173:} The variation in our measurements of the odd isotope transition frequencies show comparatively larger variation when we use different day-to-day alignments and laser configurations. The expected shifts due to hyperfine interaction should be zero because we are measuring at the ``magic angle'' of $\theta_{\rm L}=54.73^{\circ}$. However, residual polarization errors could introduce an additional uncertainty of $\pm 0.10$ MHz.

Some of the transitions associated with these isotopes are blended with other transitions, as shown in Fig. \ref{fig:ybscan}. In the case of the $^1S_0 (F=5/2) \rightarrow ^1P_1 (F^{\prime}=3/2)$ transition, the blending is significant enough to prevent our reliably extracting the transition data. The other transitions are well-enough isolated for good fitting, yet we see 1$\sigma$ variations as large 0.3 MHz. The uncertainties for these transitions in Table \ref{tab:isotope} adds the observed statistical variation with the polarization uncertainty ($\pm 0.10$ MHz) and the metrology uncertainty ($\pm 0.08$ MHz) in quadrature.

Our data disagree with the data of Ref. \cite{das2005} for the transitions Yb-173 (F=5/2), Yb-173 (F=7/2), Yb-171 (F=3/2), and Yb-171 (F=1/2) transitions by 2.64, 1.76, 2.75, and -0.02 MHz, respectively. As mentioned previously, this is a level similar to the variation seen in comparisons with other measurements from this group with frequency comb measurements. In addition, the influence of laser polarization as discussed in Sec. \ref{sec:hyperfine} was not considered in the measurements of Ref. \cite{das2005}. Using the hyperfine Hamiltonian of Ref. \cite{RevModPhys.49.31} and the data from Table \ref{tab:isotope}, we calculate the Yb-171 hyperfine coefficient to be $A = -212.33 \pm 0.30$ MHz. A comparison with other values from the literature is given in Table \ref{tab:a}. Because we do not adequately resolve the Yb-173 (F=3/2) transition, we cannot report hyperfine coefficients for that isotope.

\begin{table}
\caption{\label{tab:a} A comparison of the hyperfine $A$ coefficient for Yb-171. The column $\Delta A$ shows the difference between this work and previous determinations. The column $\Delta A/\sigma$ shows the magnitude of the difference in units of the combined uncertainties.
}
\begin{tabular}{lrcc}
$A$ (MHz) & $\Delta A$ (MHz) &  $\Delta A / \sigma$ &Ref. (year) \\ \hline \hline
$-212.33  \pm 0.30$  & $\cdots$ & $\cdots$  & This work \\
$-214.173 \pm 0.053$ & 1.84   & 6.0 & \cite{das2005} (2005) \\
$-213.08  \pm 0.47$  & 0.75   & 1.3 & \cite{banerjee2003} (2003) \\
$-211.9   \pm 3.1$   & -0.43  & 0.1 & \cite{loftus2001} (2001) \\
$-201.2   \pm 2.8$   & -11.1  & 3.9 & \cite{Deilamian:93} (1993) \\
$-213     \pm 10$    & 0.67   & 0.1 & \cite{Berends:92} (1992) \\
$-211.0   \pm 1.0$   & -1.33  & 1.3 & \cite{Liening1985} (1985) \\
$-213.4   \pm 3.0$   & 1.07   & 0.4 & \cite{Grundevik1979} (1977) \\
$-206.0   \pm 1.6$   & -6.33  & 3.9 & \cite{PhysRev.178.18} (1969) \\
$-211.0   \pm 1.0$   & -1.33  & 1.0 &\cite{chaiko1966} (1966) \\ \hline
\end{tabular}
\end{table}

\subsection{Absolute frequency of the Yb-174 transition}

Our spectroscopy method measures the frequency interval between our reference laser and the scanning laser. As shown in Table \ref{tab:transitions}, the method is accurate at the $\pm 0.08$ MHz level for intervals as large as 30 THz.

\begin{figure}
  \centerline{\includegraphics[width=\columnwidth]{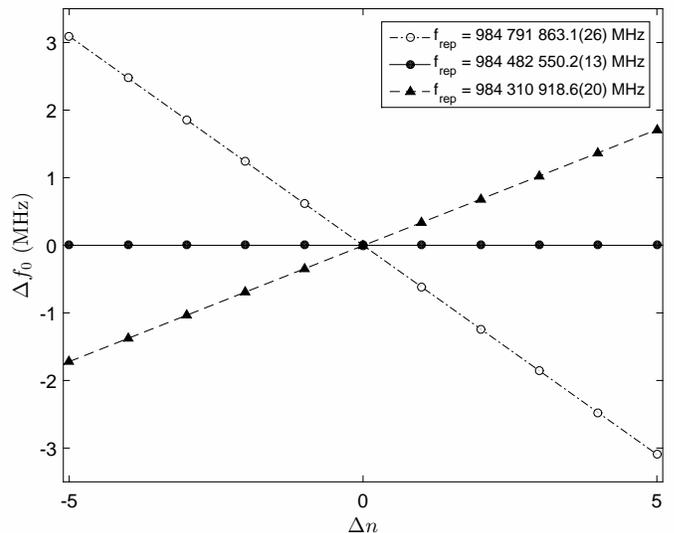}}
  \caption{\label{fig:repRateScan}
  A plot of the difference in the measured Yb-174 transition frequency with changes in the mode number $n$ for three different frequency comb repetition rates. A change in mode by $\Delta n = 1$ changes the measured frequency by nearly 2 GHz. To aid in visualization, we plot $\Delta f_0$, which is the frequency calculated using Eqn. (\ref{eqn:reprate}) with the values calculated using $f_{\rm rep} = 984.4825502$ MHz subtracted off. The uncertainties in these measurements are $\pm 0.04$ MHz. These data show that we have correctly identified the mode number $n$ in Eqs. (\ref{eqn:df}), (\ref{eqn:fscan}) and (\ref{eqn:reprate}).
  }
\end{figure}

As shown in Eqs. (\ref{eqn:df}) and (\ref{eqn:fscan}), the measured Yb transition frequency depends critically on choosing the correct value of the mode number, $n$. The wrong value would lead to a systematic shift in the Yb transition of nearly 2 GHz, equivalent to twice the value of $f_{\rm rep}$ because the scanning laser is frequency doubled. Our specified accuracy of our wavemeter is 600 MHz, which should allow accurate determination of $n$. However, we check this by measuring the Yb-174 transition frequency using different values of $f_{\rm rep}$. Different values of the repetition rate correspond to different values of $n$. But different values of $f_{\rm rep}$ and $n$ should result in the same calculated laser frequency. We therefore rewrite Eqs. (\ref{eqn:df}) and (\ref{eqn:fscan}) to obtain an expression for the frequency of the scanning laser as a function of the change in the mode number $\Delta n = n - n_0$,
\begin{equation}
  f_{\rm SL} (\Delta n) = f_{\rm Rb} - [(n_0 + \Delta n) f_{\rm rep} + f_1 + f_2],
  \label{eqn:reprate}
\end{equation}
where $n_0$ is the mode number determined using the wavemeter. We lock the frequency of the scanning laser to the center of the Yb-174 transition and measure that laser frequency using different values of $f_{\rm rep}$. The result is shown in Fig. \ref{fig:repRateScan}. The convergence of the Yb-174 transition frequency at $\Delta n = 0$ unambiguously shows that the mode number $n$ is correct.

The first-order Doppler shift contributes negligibly to the uncertainties in our measurements. The laser beam is aligned to cross the atomic beam at a right angle and the laser beam is retroreflected with an angular error of $\pm 0.2$ mRad. Given the atom velocity $v = \sqrt{3k_BT/m} = 340$ m/s and the retroreflected geometry, we expect this alignment error to result in half the Doppler shift, or $\Delta f = \pm \textstyle{\frac{1}{2}} v_{\perp}/\lambda = \pm 0.09$ MHz. A perfectly retro-reflected laser beam would result in zero shift. The retro-reflected laser beam is attenuated slightly due to absorption in the anti-reflection coated windows. Using window transmission measurements, we calculate that the retro-reflected laser beam intensity is 90\% of the incident laser beam. We verify that absorption in the laser beam due to the atomic beam is negligible. We numerically model the influence of the attenuated retro-reflected laser beam by adding two Lorentzian line profiles, one shifted up by 0.09 MHz with an amplitude of 1.0, the other shifted down by 0.09 MHz with an amplitude of 0.9, adding random noise comparable to what is seen in the experiment, and then fitting the simulated data to find the line center. We find that the fitted line center is shifted by 0.006 MHz. Because this is well below other systematic and statistical errors in our experiment, we neglect this effect in our overall uncertainty summary (see Table \ref{tab:uncertainties}).

Even though we have zeroed the magnetic field at the center of the chamber, We see an alignment-dependent shift in the measured transition frequency that is consistent with a gradient in the magnetic field of approximately 0.7 G/cm. We probe this by deliberately translating the 399 nm laser beam relative to the center of the chamber by several mm axially and transverse to the atomic beam. This shifts the apparent transition frequency by $\pm 0.30$ MHz. This gradient in the field, combined with optical pumping, may also explain the somewhat larger variation observed in the shifts of the odd isotopes in Table \ref{tab:isotope}. A summary of the uncertainties in our absolute transition frequency are listed in Table \ref{tab:uncertainties}.

\begin{table}
\caption{Summary of uncertainties in this work for the 399 nm transition in Yb-174. The laser metrology errors are twice the value shown in Table \ref{tab:transitions} because the Yb laser is frequency doubled.
\label{tab:uncertainties}}
\begin{tabular}{lc}
Source & Uncertainty ($\pm$MHz) \\ \hline \hline
Laser metrology      & 0.08 \\
Zeeman effect        & 0.05 \\
first-order Doppler shift & 0.09 \\
Statistical fitting  & 0.03 \\
$\nabla \vec{\mathbf{B}}$ & 0.30 \\ \hline
Total                & 0.33 \\ \hline
\end{tabular}
\end{table}

Our frequency for the Yb-174 $^1S_0 - ^1P_1$ transition frequency at 399 nm is
\begin{equation}
  f_0 =751\;526\;533.49 (33)~\mbox{MHz}.
  \label{eqn:absolute}
\end{equation}
This value is compared with the three previous laser-based measurements from the literature in Table \ref{tab:absolute}. The values in the literature report measurements for different isotopes. Using our frequency for Yb-174 in Eq. \ref{eqn:absolute} and our isotope shift data in Table \ref{tab:isotope}, we can compare with these different reported values. The Yb-176 transition reported in Ref. \cite{nizamani2010} has an uncertainty of 60 MHz and agrees with our value to within 125 MHz. The Yb-171 (F=3/2) transition reported in Ref. \cite{Enomoto2016} has an uncertainty of 100 MHz. It also agrees well with our value to 111 MHz. However, the value from Ref. \cite{das2005} is significantly different. Their 545 MHz variation from our value is similar to the discrepancy observed when comparing this group's value for K-39 with frequency comb measurements, as discussed in Sec. \ref{sec:accuracy}.

\begin{table}
\caption{\label{tab:absolute} A comparison of absolute transition frequencies in Yb. The number in parenthesis in column 2 is the 1$\sigma$ uncertainty in the last digits of the measurement.}
\begin{tabular}{lll}
Transition & $f_0$ (MHz) & Ref. \\ \hline
Yb-174     & 751 526 533.49(33)  & This work \\
		   & 751 525 987.761(60) & \cite{das2005} \\
		   & 751 526 650(60)	 & \cite{nizamani2010}\\
Yb-171 (F=3/2)& 751 527 368.68(39) & This work \\
              & 751 527 480(100)   & \cite{Enomoto2016} \\
\hline \hline
\end{tabular}
\end{table}

\section{Conclusion}

We present a measurement of the Yb $6s^2~^1S_0 - 6s6p~^1P_1$ transition at 399 nm and compare to values from the literature. Our frequency comb measurements agree well with the measurements of Refs. \cite{nizamani2010} and \cite{Enomoto2016} but disagree with the measurements of Ref. \cite{das2005}. We show that discrepancies between other frequency measurements by this group and the more accurate values of frequency comb based measurements are similar to the discrepancy observed here. We also show that hyperfine effects shift the apparent transition frequency in the odd-isotopes, an effect that is not addressed in previous measurements in Yb or measured in any other divalent atom.

\section{Acknowledgements}

This research was supported in part by the National Science Foundation (NSF) Grants PHY-1404488 and PHY-1500376.

\section{Note added in proof:}
After this paper was accepted for publication, the authors became aware of a publication by Wen-Li and Xin-Ye reporting isotope shifts and hyperfine constants in Yb. The reference is Chinese Physics B \bf{19}, 123202 (2010), \verb+http://iopscience.iop.org/1674-1056/19/12/123202)+.

\end{document}